\documentclass[11pt]{article}
\usepackage{amssymb}
\def\hybrid{\topmargin 0pt      \oddsidemargin 0pt
        \headheight 0pt \headsep 0pt
        \voffset=-0.5cm
        \hoffset=-0.25in
        \textwidth 6.75in
        \textheight 9.5in       % A4 paper
        \marginparwidth 0.0in
        \parskip 5pt plus 1pt   \jot = 1.5ex}
\catcode`\@=11
\def\marginnote#1{}

\newcount\hour
\newcount\minute
\newtoks\amorpm
\hour=\time\divide\hour by60 \minute=\time{\multiply\hour by60
\global\advance\minute by-\hour}
\edef\standardtime{{\ifnum\hour<12 \global\amorpm={am}%
        \else\global\amorpm={pm}\advance\hour by-12 \fi
        \ifnum\hour=0 \hour=12 \fi
        \number\hour:\ifnum\minute<10 0\fi\number\minute\the\amorpm}}
\edef\militarytime{\number\hour:\ifnum\minute<10 0\fi\number\minute}

\def\draftlabel#1{{\@bsphack\if@filesw {\let\thepage\relax
   \xdef\@gtempa{\write\@auxout{\string
      \newlabel{#1}{{\@currentlabel}{\thepage}}}}}\@gtempa
   \if@nobreak \ifvmode\nobreak\fi\fi\fi\@esphack}
        \gdef\@eqnlabel{#1}}
\def\@eqnlabel{}
\def\@vacuum{}
\def\draftmarginnote#1{\marginpar{\raggedright\scriptsize\tt#1}}
\def\draftlabel#1{{\@bsphack\if@filesw {\let\thepage\relax
   \xdef\@gtempa{\write\@auxout{\string
      \newlabel{#1}{{\@currentlabel}{\thepage}}}}}\@gtempa
   \if@nobreak \ifvmode\nobreak\fi\fi\fi\@esphack}
        \gdef\@eqnlabel{#1}}
\def\@eqnlabel{}
\def\@vacuum{}
\def\draftmarginnote#1{\marginpar{\raggedright\scriptsize\tt#1}}

\def\draft{\oddsidemargin -.5truein
        \def\@oddfoot{\sl preliminary draft \hfil
        \rm\thepage\hfil\sl\today\quad\militarytime}
        \let\@evenfoot\@oddfoot \overfullrule 3pt
        \let\label=\draftlabel
        \let\marginnote=\draftmarginnote
   \def\@eqnnum{(\theequation)\rlap{\kern\marginparsep\tt\@eqnlabel}%
\global\let\@eqnlabel\@vacuum}  }

\def\numberbysection{\@addtoreset{equation}{section}
        \def\theequation{\thesection.\arabic{equation}}}

\def\underline#1{\relax\ifmmode\@@underline#1\else
        $\@@underline{\hbox{#1}}$\relax\fi}

\def\titlepage{\@restonecolfalse\if@twocolumn\@restonecoltrue\onecolumn
     \else \newpage \fi \thispagestyle{empty}\c@page\z@
        \def\thefootnote{\fnsymbol{footnote}} }

\def\endtitlepage{\if@restonecol\twocolumn \else  \fi
        \def\thefootnote{\arabic{footnote}}
        \setcounter{footnote}{0}}  %\c@footnote\z@ }
\draft

\numberbysection \hybrid

\newcounter{mo}

\newcommand{\tr}{{\rm tr}}
\newcommand{\ti}[1]{\tilde{#1}}

\newcommand{\mH}{{\mathcal H}}

\newcommand{\vf}{\varphi}
\newcommand{\al}{\alpha}
\newcommand{\be}{\beta}
\newcommand{\ga}{\gamma}
\newcommand{\om}{\omega}
\newcommand{\vth}{\vartheta}

\newcommand{\mC}{\mathbb C}
\newcommand{\mZ}{\mathbb Z}

\newcommand{\g}{{\tilde g}}

\newtheorem{predl}{Proposition}[section]

\def\beq{\begin{equation}}
\def\eq{\end{equation}}
\def\p{\partial}

\newcommand{\mat}[4]{\left(\begin{array}{cc}{#1}&{#2}\\ \ \\{#3}&{#4}
\end{array}\right)}

\def\res{\mathop{\hbox{Res}}\limits}

\begin{document}

\setcounter{page}{1}

%\date{}
%\date{}
%\vspace{50mm}

\begin{flushright}
 ITEP-TH-21/20\\
\end{flushright}
\vspace{0mm}

\begin{center}
\vspace{10mm}
 {\LARGE{Field theory generalizations of two-body Calogero-Moser }}
 \\ \vspace{4mm} {\LARGE{models in the form of Landau-Lifshitz equations}}
% {\Large{ Explicit change of variables between Landau-Lifshitz equations}}
%\\ \vspace{4mm} {\Large{ and 1+1 Calogero-Moser field theories}}
%
\vspace{12mm}

 {\Large {K. Atalikov}\,\footnote{Institute for Theoretical and Experimental Physics of NRC ''Kurchatov Institute'',
 B.Cheremushkinskaya 25, Moscow 117218, Russia;
  e-mail: kantemir.atalikov@yandex.ru.}
 \quad\quad\quad
 {A. Zotov}\,\footnote{Steklov Mathematical Institute of Russian
Academy of Sciences, Gubkina str. 8, 119991, Moscow, Russia;
 Institute for Theoretical and Experimental Physics of NRC ''Kurchatov Institute'',
 B.Cheremushkinskaya 25, Moscow 117218, Russia; National Research
 University Higher School of Economics, Russian Federation;
%  Moscow
% Institute of Physics and Technology, Inststitutskii per.  9,
% Dolgoprudny, Moscow region, 141700, Russia;
 e-mail: zotov@mi-ras.ru.}
 }

\end{center}

%\vspace{2mm}
%\begin{center}\footnotesize{{\rm E-mails:}{\rm\
% kantemir.atalikov@yandex.ru\ zotov@mi-ras.ru}}\end{center}
%
\vspace{5mm}

 \begin{abstract}
 We give detailed description for continuous version of the classical IRF-Vertex relation, where on the IRF
 side we deal with the Calogero-Moser-Sutherland models. Our study is based on
 constructing modifications of the Higgs bundles of infinite rank over elliptic curve and its degenerations.
 In this way the previously predicted
 gauge equivalence between L-A pairs of the Landau-Lifshitz type equations and
 1+1 field theory generalization of the Calogero-Moser-Sutherland models is described. In this paper the ${\rm sl}_2$ case is studied.
  Explicit changes of variables
 are obtained between the rational, trigonometric and elliptic models.
 \end{abstract}

\bigskip

\bigskip

%\vskip30mm

%\newpage

%\small{
\tableofcontents
%}
%\newpage

%%%%%%%%%%%%%%%%%%%%%%%%%%%%%%%%%%%%%%%%%%%%%%%%%%%%%%%%%%%%%%%%%%%%%%%%%%%%%%%%%%%%%%%%%%%%%%%%%%%%%%
%%%%%%%%%%%%%%%%%%%%%%%%%%%%%%%%%%%%%%%%%%%%%%%%%%%%%%%%%%%%%%%%%%%%%%%%%%%%%%%%%%%%%%%%%%%%%%%%%%%%%%

\section{Introduction and summary}
\setcounter{equation}{0}

We begin with a brief review of the gauge equivalence between the 2-body Calogero-Moser model and the Euler top.
Then we proceed to the level of 1+1 filed generalizations based on the usage of
L-A pairs with spectral parameter on elliptic curve and satisfying the Zakharov-Shabat equation \cite{ZaSh,Kr0,Skl}.
 Our aim is to show that the gauge equivalence
holds true at this level as well and provides explicit change of variables.

\paragraph{Two-body Calogero-Moser model.}
In this paper we deal with the Calogero-Moser model \cite{Calogero}. Its elliptic version for the 2-body model (in the center of mass frame) is described by the
Hamiltonian
 \beq\label{a111}
 \begin{array}{c}
   \displaystyle{
 H_{0}^{\rm CM}=\frac{p^2}{2} - \frac{\nu}{8} ^2\wp(q)\,,
 }
 \end{array}
  \eq
and the canonical Poisson bracket
 \beq\label{a1110}
 \begin{array}{c}
   \displaystyle{
 \{p,q\}=1\,.
 }
 \end{array}
  \eq
The potential is given by the
Weierstrass elliptic function (\ref{A4}).
It is the simplest example of the $N$-body model. All variables are $\mC$-valued including the coupling constant
$\nu$ and the elliptic moduli $\tau$, so that one can easily change signs if needed. Although the model (\ref{a111}) describes a single
degree of freedom (and therefore, it is Liouville integrable) we use the Lax pair with spectral parameter $z$ given
by the pair of matrices \cite{Kr}:
 \beq\label{a107}
 \begin{array}{c}
   \displaystyle{
 L^{\mathrm{CM}}(z)=\left(\begin{array}{cc}
p & \displaystyle{\frac{\nu}{2}\, \phi(-z, q) }\\
\displaystyle{ \frac{\nu}{2}\, \phi(-z, -q) } & -p
\end{array}\right),
 }
 \quad
  \displaystyle{
 M^{\mathrm{CM}}(z)=\frac{1}{4}\left(
\begin{array}{cc}
 0 & \nu  f(-z,q) \\
  \nu  f(-z,-q) & 0 \\
\end{array}
\right),
 }
 \end{array}
  \eq
  where the functions $\phi(z, q)$ and $f(z, q)$ are given in (\ref{A5}), (\ref{A61}).
The Lax equation
 \beq\label{a108}
 \begin{array}{c}
   \displaystyle{
 \dot{L}(z)=\{H, L(z)\}=[L(z), {M}(z)]
 }
 \end{array}
  \eq
for (\ref{a111})-(\ref{a107}) holds true identically in $z$ on the equations of motion
 \beq\label{a113}
 \begin{array}{c}
   \displaystyle{
 \dot{p}=\frac{\nu ^2}{8} \wp'(q), ~~~ \dot{q}=p\,,
 }
 \end{array}
  \eq
generated by the Hamiltonian (\ref{a111}) and the Poisson bracket (\ref{a1110}). The Hamiltonian
(\ref{a111}) appears from the generating function of conservation laws:
 \beq\label{a110}
 \begin{array}{c}
   \displaystyle{
 \frac{1}{4}\,\tr\Big(L^{\rm CM}(z)\Big)^2=H_{0}^{\rm CM} + \frac{\nu ^2}{8}\,\wp(z)\,.
 }
 \end{array}
  \eq

\paragraph{Complexified Euler top in ${\mC}^3$.} It is defined by the Hamiltonian
 \beq\label{a1161}
 \begin{array}{c}
   \displaystyle{
H_0^{\rm top}= -\frac{1}{2}\,\sum\limits_{\al=1}^3 S_\al^2\wp(\om_\al)
 }
 \end{array}
  \eq
  and the Poisson-Lie structure on ${\rm sl}^*(2,\mC)$ (complexification of ${\rm su^*(2)}$)
 \beq\label{a135}
 \begin{array}{c}
   \displaystyle{
 \left\{S_{\alpha}, S_{\beta}\right\}=- \sqrt{-1} \varepsilon_{\alpha \beta \gamma} S_{\gamma}\,.
 }
 \end{array}
  \eq
The variables $S_\al$ are naturally arranged into the traceless $2\times 2$ matrix
$S=\sum\limits_{\al=1}^3\sigma_\al S_\al$ in the Pauli matrices basis.
Equations of motion  $ \dot{S}=\{H_0^{\rm top}, S\}$ then takes the form
 \beq\label{a123}
 \begin{array}{c}
   \displaystyle{
 \dot{S}=[J(S), S]\qquad J(S)=\sum\limits_{\al=1}^3 S_\al J_\al\sigma_\al\,,\quad J_\al=-\frac{1}{2}\,\wp(\om_\al)\,.
 }
 \end{array}
  \eq
Notice again
that all variables are complex. The vector $(S_1,S_2,S_3)$ is a complexification of the angular momentum vector
of a rigid body (the Euler top). The constants $J_\al$ are components of the inverse inertia tensor in principle axes.
In (\ref{a123}) all $J_\al$ depend on a single parameter -- elliptic moduli $\tau$. In order to have a set of free parameters $J_\al$ one may
multiply $H_0^{\rm top}$ by an arbitrary constant and also shift it by an expression proportional to the Casimir function
$C_2= \frac{1}{2}\,\sum\limits_{\al=1}^3 S_\al^2$ of (\ref{a135}). Equations of motion (\ref{a123}) for the Hamiltonian (\ref{a1161}) admit
the Lax representation (\ref{a108}). The Lax pair \cite{Skl} is
 \beq\label{a115}
 \begin{array}{c}
   \displaystyle{
 %L^{\mathrm{top}}(z)=
 L^{\mathrm{top}}(z,S)=\sum\limits_{\al=1}^3 S_\al\vf_\al(z)\sigma_\al\,,
 \qquad
 %M^{\mathrm{top}}(z)=
 M^{\mathrm{top}}(z,S)=-\frac{1}{2}\,\sum\limits_{\al=1}^3 S_\al f_\al(z)\sigma_\al\,.
 }
 \end{array}
  \eq
Similarly to (\ref{a110}) we have
 \beq\label{a116}
 \begin{array}{c}
   \displaystyle{
 \frac{1}{4}\,\tr\Big(L^{\mathrm{top}}(z)^2\Big)= \frac{1}{2}\,\sum\limits_{\al=1}^3 S_\al^2\vf_\al(z)^2
 =H_0^{\rm top}+C_2\wp(z)\,,
 }
  \\ \ \\
   \displaystyle{
 H_0^{\rm top}=\frac{1}{2}\,\tr(SJ(S))\,,\qquad C_2=\frac14\,\tr(S^2)\,.
 }
 \end{array}
  \eq

\paragraph{Gauge equivalence.} The above models were shown to be gauge equivalent in ${\rm gl}_N$ case \cite{LOZ}.
The ${\rm sl}_2$  case is the simplest example
of the Symplectic Hecke correspondence. Introduce the matrix\footnote{The gauge equivalence can be also viewed as the classical version of the IRF-Vertex relation. The matrix of type (\ref{a133}) is known from studies
of the 8-vertex quantum models \cite{Baxter2}.}
 \beq\label{a133}
 \begin{array}{c}
   \displaystyle{
 g(z)=\left(\begin{array}{cc}
\theta_{00}(z+ q\,|\,2 \tau) & -\theta_{00}(z- q\,|\,2 \tau) \\
-\theta_{10}(z+ q\,|\,2 \tau) & \theta_{10}(z- q\,|\,2 \tau)
\end{array}\right)\,.
 }
 \end{array}
  \eq
The gauge equivalence
 \beq\label{a130}
 \begin{array}{c}
   \displaystyle{
 L^{\rm top}(z)=g(z){L}^{\rm CM}(z){g}^{-1}(z)
 }
 \end{array}
  \eq
  holds true identically in $z$ and provides explicit change of variables $S_\al=S_\al(p,q,\nu)$:
 \beq\label{a134}
 \left\{\begin{array}{l}
   \displaystyle{
   S_{1}(p,q,\nu) =p \frac{\theta_{01}(0)}{\vartheta^{\prime}(0)} \frac{\theta_{01}(q)}{\vartheta(q)}+\frac{\nu}{2} \frac{\theta_{01}^{2}(0)}{\theta_{00}(0) \theta_{10}(0)} \frac{\theta_{00}(q) \theta_{10}(q)}{\vartheta^{2}(q)}\,,
   }
    \\ \ \\
    \displaystyle{
   S_{2}(p,q,\nu) =p \frac{\sqrt{-1}\theta_{00}(0)}{ \vartheta^{\prime}(0)} \frac{\theta_{00}(q)}{\vartheta(q)}
   +\frac{\nu}{2} \frac{\sqrt{-1}\theta_{00}^{2}(0)}{ \theta_{10}(0) \theta_{01}(0)}
    \frac{\theta_{10}(q) \theta_{01}(q)}{\vartheta^{2}(q)}\,,
   }
    \\ \ \\
   \displaystyle{
    S_{3}(p,q,\nu) =p \frac{\theta_{10}(0)}{\vartheta^{\prime}(0)} \frac{\theta_{10}(q)}{\vartheta(q)}+\frac{\nu}{2} \frac{\theta_{10}^{2}(0)}{\theta_{00}(0) \theta_{01}(0)} \frac{\theta_{00}(q) \theta_{01}(q)}{\vartheta^{2}(q)}\,.
 }
 \end{array}
 \right.
  \eq

  The map $(p,q)\mapsto (S_1,S_2,S_3)$ is symplectic, i.e. the Poisson brackets between the functions $S_\al(p,q,\nu)$ evaluated through
  the canonical brackets (\ref{a1110}) provide the linear Poisson-Lie structure (\ref{a135}). The relation (\ref{a130})
  implies equality of the r.h.s. of  (\ref{a110}) and (\ref{a116}), that is the Casimir function
  (\ref{a116}) is fixed as $C_2=\nu^2/8$. In this way the symplectic leaf of (\ref{a135})
  is fixed through the value of the Calogero-Moser coupling constant. It means that the matrix $S(p,q,\nu)$ has fixed eigenvalues $\{\lambda,-\lambda\}$ with $\lambda=\nu/2$.
  Using identity (\ref{A29}) let us put (\ref{a134}) in a more convenient and compact form:
 \beq\label{a1341}
 \begin{array}{c}
   \displaystyle{
   S_{\al}(p,q,\nu) = c_\al(\tau) \Big(p-\frac{\nu}{2}\,\p_q\Big)\vf_\al(q)\,,
   }
 \end{array}
  \eq
 \beq\label{a1342}
 \begin{array}{c}
   \displaystyle{
    c_1(\tau)=\Big(\frac{\theta_{01}(0)}{\vartheta^{\prime}(0)}\Big)^2\,,\quad
    c_2(\tau)=\sqrt{-1}\Big( \frac{\theta_{00}(0)}{ \vartheta^{\prime}(0)} \Big)^2\,,\quad
    c_3(\tau)=\Big( \frac{\theta_{10}(0)}{\vartheta^{\prime}(0)} \Big)^2\,.
   }
 \end{array}
  \eq

\paragraph{Landau-Lifshitz equation.} The field generalization for the Euler top (\ref{a123}) is the
Landau-Lifshitz equation \cite{LL,Skl}:
 \beq\label{a233}
 \begin{array}{c}
   \displaystyle{
 \partial_{t} S=[J(S), S]-\alpha_0\left[S, S_{x x}\right]\,,\quad S_{xx}=\p_x^2S\,,
 }
 \end{array}
  \eq
 where $J(S)$ is from (\ref{a123}) and
 \beq\label{a2338}
 \begin{array}{c}
   \displaystyle{
\alpha_0=k^2/(8\lambda^2)
 }
 \end{array}
  \eq
is a constant parameter. Here $\lambda\in\mC$ (also constant) is an eigenvalue of the matrix $S$ -- the norm of the vector $(S_1,S_2,S_3)$, and $k$ is a constant coefficient behind $\p_x$. The limit $k\rightarrow0$ provides transition to the
 classical mechanics of the Euler top (\ref{a123}). The dynamical variables
 %
 % !!! $\al=??$ $\al$ plokho - drugoe oboznach
 %
 % where
 $S_\al=S_\al(t,x)$ are fields on 1+1 spacetime. In the real case $(S_1,S_2,S_3)$
  is a magnetization (normalized) vector in ${\mathbb R}^3$. We deal with a complexified version
  of the Landau-Lifshitz equation similarly to complexification of the Euler top.
  At the same time $x$ is coordinate on a circle: $x\in {\mathbb S}^1$. The
  Poisson-Lie structure is the one on the loop Lie coalgebra $L^*({\rm sl}(2,\mC))$:
 \beq\label{a232}
 \begin{array}{c}
   \displaystyle{
 \left\{S_{\alpha}(x), S_{\beta}(y)\right\}=-\sqrt{-1} \varepsilon_{\alpha \beta \gamma} S_{\gamma}(x)\delta (x-y)\,,
 }
 \end{array}
  \eq
The fields are periodic functions on a circle with coordinate $x$. Equation (\ref{a233}) is generated by (\ref{a232}) and the Hamiltonian
 \beq\label{a2327}
 \begin{array}{c}
   \displaystyle{
  \mH^{\rm LL}=\frac{1}{2}\oint dx \Big( \tr(SJ(S))-\al_0\tr(S_x^2) \Big)\,.
 }
 \end{array}
  \eq
 It was shown in \cite{Skl} that the Landau-Lifshitz equation (\ref{a233}) is represented in the Zakharov-Shabat (or zero curvature) form \cite{ZaSh}
 \beq\label{a11}
 \begin{array}{c}
   \displaystyle{
    \partial_{t} U(z)+k \partial_{x} V(z)=[U(z), V(z)]
     }
 \end{array}
   \eq
identically in $z$ with
 \beq\label{a1196}
 \begin{array}{c}
   \displaystyle{
    U^{\rm LL}(z)=L^{\rm top}(z,S)=\sum\limits_{\al=1}^3 S_\al\vf_\al(z)\sigma_\al\,,
     }
 \end{array}
   \eq
 \beq\label{a1197}
 \begin{array}{c}
   \displaystyle{
    V^{\rm LL}(z)=\frac{1}{2}\,\Big( E_1(z)L^{\rm top}(z,S)+2M^{\rm top}(z,S)-L^{\rm top}(z,v) \Big)=
    }
    \\ \ \\
    \displaystyle{
    =\frac{1}{2}\sum\limits_{\al=1}^3 \Big(S_\al\vf_\be(z)\vf_\ga(z)-v_\al\vf_\al(z)\Big)\sigma_\al\,,
     }
 \end{array}
   \eq
where the matrix $v$
 \beq\label{a1198}
 \begin{array}{c}
   \displaystyle{
    v=\sum\limits_{\al=1}^3v_\al\sigma_\al=-\frac{k}{4\lambda^2}[S,S_x]\in{\rm Mat}_2\,,
     }
 \end{array}
   \eq
is a solution of condition $kS_x=[v,S]$ solved by the usage of the characteristic equation $S^2=\lambda^21_{2\times 2}$ and its derivative: $SS_x+S_xS=0$.

\paragraph{1+1 field generalizations of Calogero-Moser system.} In this case the momentum and coordinate become fields
on ${\mathbb S}^1$ and the canonical Poisson bracket (\ref{a1110}) turns into
 \beq\label{a222}
 \begin{array}{c}
   \displaystyle{
 \left\{p(x), q(y)\right\}= \delta(x-y)\,.
 }
 \end{array}
  \eq
  It was shown in \cite{Kr2} and  \cite{LOZ}  that the two-body Hamiltonian is generalized as follows:
 \beq\label{a220}
 \begin{array}{c}
   \displaystyle{
 \mH^{\rm CM}= \frac{1}{2}\oint dx \Big( {{p}^{2}}\left( 1-\frac{{{k}^{2}}q_{x}^{2}}{{{c}^{2}}} \right)+\frac{(3{{k}^{2}}q_{x}^{2}-{{c}^{2}})}{4}\wp(q)-\frac{{{k}^{4}}{{q}^{2}_{xx}}}{4(c^2-k^2q_x^2)}\Big)\,,
 }
 \end{array}
  \eq
  where $c$ is a constant. Equations of motion takes the form:
 \beq\label{a221}
 \begin{array}{c}
   \displaystyle{
 q_t={p}\left( 1-\frac{{{k}^{2}}q_{x}^{2}}{{{c}^{2}}} \right)\,,
 }
 \\ \ \\
    \displaystyle{
  p_t=-\frac{ k^2}{ c^2} \partial_{x}\left(p^{2} q_{x}\right)-\frac{(3{{k}^{2}}q_{x}^{2}-{{c}^{2}})}{8}\wp'(q)+\frac{3 k^2}{4} \partial_{x}\left( q_{x} \wp(q) \right)+\frac{k^4}{4} \partial_{x}\left(\frac{q_{x x x} {\tilde\nu}-{\tilde\nu}_{x} q_{x x}}{{\tilde\nu}^{3}}\right)\,,
  }
  \end{array}
   \eq
where
%$\tilde\nu=\sqrt{c^2-k^2q_x^2}$.
  %
 \beq\label{a2216}
 \begin{array}{c}
   \displaystyle{
\tilde\nu=\sqrt{c^2-k^2q_x^2}\,,\quad c={\rm const}\,.
  }
  \end{array}
   \eq
%   %
%
These equations are represented in the Zakharov-Shabat form (\ref{a11}) with
%L-A pair
   %
 \beq\label{a2201}
 \begin{array}{c}
   \displaystyle{
 U^{\rm CM}(z)=\frac12\mat{2p-kq_xE_1(z)}{\tilde\nu\phi(-z,q)}{\tilde\nu\phi(-z,-q)}{-2p+kq_xE_1(z)}
 }
 \end{array}
  \eq
  and traceless $V^{\rm CM}(z)$ with  matrix elements
  %($V_{22}^{\rm CM}(z)=-V_{11}^{\rm CM}(z)$):
     %
 \beq\label{a2202}
 \begin{array}{c}
   \displaystyle{
   V_{11}^{\rm CM}(z)=\frac14\left({2q_tE_1(z)+\frac{4k}{c^2}p^2q_x-3kq_x\wp(q)-k^3\Big(\frac{q_{x x x} {\tilde\nu}-{\tilde\nu}_{x} q_{x x}}{{\tilde\nu}^{3}}\Big)}+kq_x(E_2(q)-E_2(z))\right)\,,
   }
 \end{array}
  \eq
 \beq\label{a2203}
 \begin{array}{c}
   \displaystyle{
   V_{22}^{\rm CM}(z)=-V_{11}^{\rm CM}(z)
   }
 \end{array}
  \eq
       and
 \beq\label{a2204}
 \begin{array}{c}
    \displaystyle{
 V_{12}^{\rm CM}(z)=\frac{\tilde\nu}{4} \left( f(-z,q)+\Big( E_1(z)+\frac{2k pq_x}{c^2}+\frac{k^2q_{xx}}{\ti\nu^2}\Big)\phi(-z,q)\right)\,,
 }
 \end{array}
  \eq
 \beq\label{a2205}
 \begin{array}{c}
     \displaystyle{
 V_{21}^{\rm CM}(z)=\frac{\tilde\nu}{4} \left( f(-z,-q)+\Big( E_1(z)+\frac{2k pq_x}{c^2}-\frac{k^2q_{xx}}{\ti\nu^2}\Big)\phi(-z,-q)\right)\,.
 }
 \end{array}
  \eq
  In the limit to the classical mechanics $k\rightarrow 0$ the constant $c$ becomes the coupling constant $\nu$ (together with $\ti\nu$):
  $\ti\nu=c=\nu$. In this limit the matrix $V$ (\ref{a2202})-(\ref{a2205}) turns into
 \beq\label{a2206}
 \begin{array}{c}
     \displaystyle{
  V^{\rm CM}(z)\stackrel{k=0}{\longrightarrow} M^{\rm CM}(z)+\frac12\,E_1(z)L^{\rm CM}(z)\,,
 }
 \end{array}
  \eq
  and the second term is not necessary in mechanics since it is cancelled out from the Lax equation (\ref{a108}).

\paragraph{Purpose of the paper.} While the Lax equations (\ref{a108}) are gauge invariant in the general case
with respect to adjoint action $L\rightarrow f Lf^{-1}$ (together with $M\rightarrow f Mf^{-1}-{\dot f}f^{-1}$), the matrix $U$ in the zero curvature equation (\ref{a11}) is a component of the connection along the loop ${\mathbb S}^1$ in the algebra $L({\rm sl}(2,\mC))$, so that
it transforms as $U\rightarrow f Uf^{-1}+kf_xf^{-1}$. Based on the gauge equivalence at the level
of classical mechanics (\ref{a130}) it is reasonable to expect a similar relation for 1+1 models:
 \beq\label{a250}
 \begin{array}{c}
   \displaystyle{
U^{\rm LL}(z)=\g U^{CM}(z) \g^{-1}+k \g_x \g^{-1}
 }
 \end{array}
  \eq
with some gauge transformation $\g$. It can be verified that the choice (\ref{a134}) $\g=g$ from mechanics does not match the field case since the r.h.s. of
(\ref{a250}) with $\g=g$ provides different from $U^{\rm LL}(z)$ dependence on $z$. We need the r.h.s. of
(\ref{a250}) to be a matrix valued function (of $z$) of the form $U^{\rm LL}(z)$ (\ref{a1196}). Then the coefficients behind
$\vf_\al(z)\sigma_\al$ could be identified with $S_\al$.

The aim of the paper is to define
the gauge transformation $\g(z)$ and find the change of variables between the Landau-Lifshitz model (\ref{a232})-(\ref{a2327})
and the 1+1 Calogero-Moser model (\ref{a222})-(\ref{a220}) in the elliptic, trigonometric and rational cases.

It must be emphasized that (as was mentioned in \cite{Kr2}) the change of variables between (\ref{a233}) and (\ref{a221})
was already found by A. Shabat in the elliptic case.
His aim was to include equation  (\ref{a221}) into the classification of the second order
nonlinear integrable equations \cite{MSh}.
%As far as we know, his answer for the change of variables was not published.
Unfortunately, we could not find these results in published papers to compare to ours.

 Our construction is based on the group-theoretical approach of \cite{LOZ}. Its advantage is that it can be generalized to
 $N$-body 1+1 Calogero-Moser model. We plan to study it in future publications. Also, our method is applicable
 to trigonometric and rational models. While the trigonometric and rational limits are simple on the side of Calogero-Moser model, they are quite difficult at the Landau-Lifshitz side. The underlying $R$-matrices (in the discrete cases)
are the 7-th and 11-th vertex $R$-matrices found in \cite{Chered}. Our approach allows to avoid these difficulties. We do not
perform the scaling limits from the elliptic model, but just choose an appropriate $\g$ and apply the gauge transformation
(\ref{a250}) to the rational and trigonometric Calogero-Moser models. This strategy yields the correct answers.

 Let us also remark that the gauge equivalence between integrable hierarchies is a widely known phenomenon.
 We mention three examples close to our case. The first one is the gauge equivalence between 1+1  Heisenberg magnet
(the real case of the Landau-Lifshitz equation (\ref{a233}) with $J(S)=0$) and the nonlinear Schrodinger equation
(in some special case) \cite{ZaTa}. Its extensions to the Landau-Lifshitz case are also known \cite{Kundu}. The canonical coordinates in the NLS equation differ from those in the Calogero-Moser model. We will clarify relation of our results to NLS
type models in our future works. The second example is the gauge transformation with the classical
$r$-matrix structure in WZNW (and Toda) theories described in \cite{Feher}.
And the third example is the already mentioned IRF-Vertex relation in the quantum statistical models \cite{Baxter2}.
%Two latter examples are indeed close to our study.
Our results can be treated as 1+1 continuous version of the classical IRF-Vertex relation %(performed for the continuous 1+1 models)
for the vertex models with elliptic Baxter's $R$-matrix and those degenerations described in \cite{Chered}.

The paper is organized as follows. In the next Section the elliptic case is studied. We explain how
to modify the definition of the gauge transformation (\ref{a133}) in order to obtain $\g$.
 Then we apply it as in (\ref{a250}) and derive the field generalization for the change of variables
 (\ref{a134})-(\ref{a1342}). In Sections \ref{sect3} and \ref{sect4} similar results
 are obtained in trigonometric and rational cases. A brief summary of results is made in the Conclusion.
 In the Appendix we collect the elliptic function identities needed for calculations in Section \ref{sect2}.

\section{Elliptic models}\label{sect2}
\setcounter{equation}{0}

We first recall how the matrix $g(z)$ was defined in the classical mechanics in (\ref{a130}), and then
explain how it should be modified in the 1+1 field case.

\subsection{Construction of the gauge transformation}
\paragraph{Mechanical case.}
Let us represent $g(z)$ (\ref{a133}) in the form:
 \beq\label{a245}
 \begin{array}{c}
   \displaystyle{
g(z)=\Xi(z)D^{-1}\,,
\quad
\Xi(z)=\left(\begin{array}{cc}
\theta_{00}(z+ q\,|\,2 \tau) & \theta_{00}(z- q\,|\,2 \tau) \\
-\theta_{10}(z+ q\,|\,2 \tau) & -\theta_{10}(z- q\,|\,2 \tau)
\end{array}\right)\,,
 \quad
D^{-1}=\left(\begin{array}{cc}
1 & 0 \\
 0 & -1
\end{array}\right)
}
 \end{array}
  \eq
  The matrix $g(z)$ is defined by the set of properties \cite{LOZ}.
 First of all $g(z)$  must properly change the transition functions
  of the Lax matrices describing their behavior on the lattice of periods. The Lax matrix
  is a section of the Higgs bundle over elliptic curve. More precisely, it is a
  section of ${\rm End}(V)$-bundle for some holomorphic vector bundle $V$ given
  by its transition functions $h_1(z)$ and $h_\tau(z)$ on the torus. So that
 \beq\label{a2001}
 \begin{array}{c}
   \displaystyle{
L^{\rm }(z+1)=h_1(z)L^{\rm }(z)h_1^{-1}(z)\,,\quad L^{\rm }(z+\tau)=h_\tau(z)L^{\rm }(z)h_\tau^{-1}(z)\,.
}
 \end{array}
  \eq
  The underlying vector bundles for $L^{\rm CM}(z)$ and $L^{\rm top}(z)$ are
   different\footnote{As is explained in \cite{LOZ} the transformation $g(z)$ changes degree of underlying vector bundle since
  it adds a zero (to sections) at $z=0$: $\det g(z)\sim\vth(z)$. Such maps are called modifications of bundles. They are locally fixed by the choice of a point ($z=0$) and a direction in the fiber. The latter is fixed by the matrix $D$ in (\ref{a245}). See below.}.
  It follows from the consistency
  of the conditions (\ref{a2001}) applied to both sides of (\ref{a130}) that
  the gauge transformation $g(z)$ should have the following properties (up to a scalar factor):
 \beq\label{a2002}
 \begin{array}{c}
   \displaystyle{
g(z+1)=h_1^{\rm top}(z)g^{\rm }(z)(h^{\rm CM}_1(z))^{-1}\,,\quad
g(z+\tau)=h_\tau^{\rm top}(z)g^{\rm }(z)(h^{\rm CM}_\tau(z))^{-1}\,.
}
 \end{array}
  \eq
  Using (\ref{A8})-(\ref{A82}) one can find the transition functions
  for the mechanical model (\ref{a107})
 \beq\label{a2003}
 \begin{array}{c}
   \displaystyle{
h^{\rm CM}_1(z)=1_{2\times 2}\,,\quad h^{\rm CM}_\tau(z)=\exp(\pi\imath q\sigma_3)\,,
}
 \end{array}
  \eq
  and for the Lax matrix (\ref{a115}) of the Euler top:
 \beq\label{a2004}
 \begin{array}{c}
   \displaystyle{
h^{\rm top}_1(z)= \sigma_3\,,\quad h^{\rm top}_\tau(z)= \exp(\pi\imath z)\sigma_1\,.
}
 \end{array}
  \eq
The matrix $\Xi(z)$ entering $g(z)$ in (\ref{a245}) satisfies (\ref{a2002}) up to a scalar constant factor (it is cancelled out in the adjoint action). Moreover, the matrix $\Xi(z)$ saves this property after multiplication from the right by any diagonal matrix since $h^{\rm CM}_1(z)$ and $h^{\rm CM}_\tau(z)$ are also diagonal.

In this way we come to $g(z)=\Xi(z)D^{-1}$ with an arbitrary diagonal matrix $D$. This freedom is fixed by the next argument.
The matrix $\Xi(z)$ is degenerated at $z=0$:
 \beq\label{a2005}
 \begin{array}{c}
   \displaystyle{
 \det \Xi(z)\stackrel{(\ref{A14})}{=}-\vth(z)\vth(q)\,.
}
 \end{array}
  \eq
Therefore, $\Xi(z)^{-1}$ has simple pole at $z=0$, and the conjugation of the Lax matrix $L^{CM}(z)$ (which also has simple pole at $z=0$) may result in appearance of the second order pole, which is absent in $L^{top}(z)$. A sufficient condition to avoid
appearance of the second order pole is as follows \cite{LOZ}: an eigenvector $\vec{u}$ of the residue matrix $\res\limits_{z=0}L^{CM}(z)$ should belong to the kernel of $g(0)$, i.e.
 \beq\label{a2006}
 \begin{array}{c}
   \displaystyle{
g(0)\vec{u}=0\quad {\rm for}\quad \vec{u}\in{\rm Eigenspace}\Big(\res\limits_{z=0}L^{CM}(z)\Big)\,.
}
 \end{array}
  \eq
  For the matrix $L^{CM}(z)$ (\ref{a107}) its residue
 \beq\label{a2007}
 \begin{array}{c}
   \displaystyle{
 \res\limits_{z=0}L^{CM}(z)=-\frac{\nu}{2}\left(\begin{array}{cc}
0 & 1 \\
 1 & 0
\end{array}\right)
}
 \end{array}
  \eq
has eigenvectors $\vec{u}_1=(1,1)^T$ and $\vec{u}_2=(1,-1)^T$. Moreover, $\Xi(0)\vec{u}_2=0$, so that we may keep the choice $g(z)=\Xi(z)$.
The second possibility is to choose the eigenvector $\vec{u}_1=D \vec{u}_2$ with $D$ from (\ref{a245}). It is more natural from the viewpoint of ${\rm gl}_N$ model since
$\res\limits_{z=0}L^{CM}(z)$ has an eigenvector $(1,..,1)^T$. By this reason we choose $\vec{u}_1$. Then, the matrix $\Xi(z)$ should be
multiplied from the right by $D^{-1}$ with $D$ from (\ref{a245}). In this case $g(z)=\Xi(z)D^{-1}$ as it is given in (\ref{a245}). It satisfies
the condition (\ref{a2007}) for $\vec{u}=\vec{u}_1$.

\paragraph{Field case.} While the matrix $U^{\rm LL}(z)$ (\ref{a1196}) has the same form as the top's Lax matrix (\ref{a115}), the matrix
 $U^{\rm CM}(z)$ (\ref{a2201}) differs from the one $L^{\rm CM}(z)$ in mechanics. Its behavior on the lattice of periods takes the following form:
 \beq\label{a2008}
 \begin{array}{c}
   \displaystyle{
 U^{\rm CM}(z+1)=U^{\rm CM}(z)\,, \quad
 U^{\rm CM}(z+\tau)=h_\tau^{\rm CM}U^{\rm CM}(z)(h_\tau^{\rm CM})^{-1}+\pi\imath kq_x \sigma_3\,.
}
 \end{array}
  \eq
  The additional term $\pi\imath kq_x \sigma_3$ comes from the function $E_1(z)$ in the diagonal elements. It is explained as follows.
  $U^{\rm CM}(z)$ is
 a connection along the loop direction:
 \beq\label{a2009}
 \begin{array}{c}
   \displaystyle{
 k\p_x-U^{\rm CM}(z+\tau)=k\p_x-h_\tau^{\rm CM}U^{\rm CM}(z)(h_\tau^{\rm CM})^{-1}-\pi\imath kq_x \sigma_3=
 }
 \\ \ \\
    \displaystyle{
 =h_\tau^{\rm CM}\Big(k\p_x-U^{\rm CM}(z)\Big)(h_\tau^{\rm CM})^{-1}\,,
}
 \end{array}
  \eq
  so that the total connection transforms in a proper manner.

  Hence, using the arguments from the previous paragraph we come to the matrix of gauge transformation
  ${\tilde g}$ from (\ref{a250}) of the same form as $g(z)$ (\ref{a245}) but with possibly different diagonal matrix
  ${\tilde D}$:
 \beq\label{a2010}
 \begin{array}{c}
   \displaystyle{
 {\tilde g}(z)=\Xi(z){\tilde D}^{-1}\,.
}
 \end{array}
  \eq
  As seen above the matrix ${\tilde g}(z)$ should satisfy direct analogue of the condition (\ref{a2006}):
 \beq\label{a2011}
 \begin{array}{c}
   \displaystyle{
{\tilde g}(0)\vec{u}=0\quad {\rm for}\quad \vec{u}\in{\rm Eigenspace}\Big(\res\limits_{z=0}U^{CM}(z)\Big)\,.
}
 \end{array}
  \eq
  Let us solve (\ref{a2011})  with the requirement that in the classical mechanics limit ($k=0$) the answer for
  ${\tilde D}$ turns into $D$ (up to a scalar constant). We have the following residue matrix
 \beq\label{a2012}
 \begin{array}{c}
   \displaystyle{
 \res\limits_{z=0}U^{CM}(z)=-\frac{1}{2}\left(\begin{array}{cc}
kq_x & \ti\nu \\
 \ti\nu & -kq_x
\end{array}\right)\,.
}
 \end{array}
  \eq
  Using the relation (\ref{a2216}) we easily find its eigenvalues to be equal to $\pm c/2$. We choose the vector
 \beq\label{a243}
 \begin{array}{c}
   \displaystyle{
 {\vec u}_1=\left( \begin{array}{cc} -\frac{k q_{x}}{2}-\frac{c}{2} \\ -\frac{{\tilde\nu}}{2} \\ \end{array} \right)
}
 \end{array}
  \eq
  corresponding to the eigenvalue $-c/2$ since this vector is proportional to $(1,1)^T$ when $k=0$. Finally, we should
  find $\ti D$ from the condition (\ref{a2011}). It leads to
 \beq\label{a247}
 \begin{array}{c}
   \displaystyle{
 {\ti D}=\rho\left( \begin{array}{cc}
\frac{1}{{\tilde\nu} } & 0 \\
0 & \frac{-1}{k{{q}_{x}}+c} \\
\end{array} \right)\,,
 }
 \end{array}
  \eq
  where $\rho$ is a normalization factor.

  In contrast to mechanics the factor $\rho$ is important in the field case due to the term
  $k{\ti g}_x{\ti g}^{-1}$ in the gauge transformation (\ref{a250}). Both $U^{\rm CM}(z)$ and $U^{\rm CM}(z)$
  are traceless matrices. Therefore, we must have $k{\ti g}_x{\ti g}^{-1}\in{\rm sl}(2,\mC)$. For this reason
  let us fix the factor $\rho$ to satisfy condition ${\ti g}\in{\rm SL}(2,\mC)$, that is $\det {\ti g}=1$ or
  \beq\label{a2471}
  \begin{array}{c}
    \displaystyle{
  \det {\ti g}(z)=\rho^{-2}\ti\nu(c+k q_{x})\vartheta(z) \vartheta(q)=1\,.
  }
  \end{array}
   \eq
   %
   %and
   %
Thus,
  \beq\label{a2472}
  \begin{array}{c}
    \displaystyle{
  \rho=\sqrt{\ti\nu(c+k q_{x})\vartheta(z) \vartheta(q)}\,,
  }
  \end{array}
   \eq
   %
   %and
   %
and we come to the following gauge transformation (\ref{a2010}) for the relation (\ref{a250}):
 \beq\label{a248}
 \begin{array}{c}
   \displaystyle{
 {\ti g}(z)=\frac{1}{\rho}\left(
\begin{array}{cc}
 \theta_{00} (z+q \,|\, 2 \tau ) {\tilde\nu}  & -\theta_{00} (q-z \,|\, 2\tau ) \left(c+k q_{x}\right) \\ \ \\
 -\theta_{10} (z+q \,|\, 2 \tau ) {\tilde\nu}  & \theta_{10} (q-z \,|\, 2 \tau ) \left(c+k q_{x}\right)
\end{array}
\right)\,.
 }
 \end{array}
  \eq

\subsection{Change of variables}

Let us formulate the main result.
\begin{predl}\label{predl1}
 The gauge transformation (\ref{a250}) with ${\ti g}(z)$ (\ref{a248}) relates the $U$-matrices of the 1+1 Calogero-Moser model (\ref{a2201})
 and the one (\ref{a1196}) of the Landau-Lifshitz model.

\noindent 1. It provides explicit change of the field
 variables $S_\al(x)=S_\al(p(x),q(x),c)$:
 \beq\label{a251}
 \left\{\begin{array}{l}
   \displaystyle{
   S_{1}(p,q,c) =\Big(p-\frac{c}{2}\,\frac{k^2q_{xx}}{c^2-k^2q_x^2}\Big) \frac{\theta_{01}(0)}{\vartheta^{\prime}(0)} \frac{\theta_{01}(q)}{\vartheta(q)}+\frac{c}{2}\, \frac{\theta_{01}^{2}(0)}{\theta_{00}(0) \theta_{10}(0)} \frac{\theta_{00}(q) \theta_{10}(q)}{\vartheta^{2}(q)}\,,
   }
    \\ \ \\
    \displaystyle{
   S_{2}(p,q,c) =\Big(p-\frac{c}{2}\,\frac{k^2q_{xx}}{c^2-k^2q_x^2}\Big) \frac{\sqrt{-1}\theta_{00}(0)}{ \vartheta^{\prime}(0)} \frac{\theta_{00}(q)}{\vartheta(q)}
   +\frac{c}{2}\, \frac{\sqrt{-1}\theta_{00}^{2}(0)}{ \theta_{10}(0) \theta_{01}(0)}
    \frac{\theta_{10}(q) \theta_{01}(q)}{\vartheta^{2}(q)}\,,
   }
    \\ \ \\
   \displaystyle{
    S_{3}(p,q,c) =\Big(p-\frac{c}{2}\,\frac{k^2q_{xx}}{c^2-k^2q_x^2}\Big) \frac{\theta_{10}(0)}{\vartheta^{\prime}(0)} \frac{\theta_{10}(q)}{\vartheta(q)}+\frac{c}{2}\, \frac{\theta_{10}^{2}(0)}{\theta_{00}(0) \theta_{01}(0)} \frac{\theta_{00}(q) \theta_{01}(q)}{\vartheta^{2}(q)}\,,
 }
 \end{array}
 \right.
  \eq
 or in a more compact form with the notations (\ref{a1342}):
 \beq\label{a2511}
 \begin{array}{c}
   \displaystyle{
   S_{\al}(p,q,c) = c_\al(\tau) \Big(p-\frac{c}{2}\,\frac{k^2q_{xx}}{c^2-k^2q_x^2}-\frac{c}{2}\,\p_q\Big)\vf_\al(q)=
   }
   \\ \ \\
    \displaystyle{
    = c_\al(\tau) \Big(p-\frac{c}{2}\,\frac{k^2q_{xx}}{c^2-k^2q_x^2}\Big)\vf_\al(q)+c_\al(\tau)\frac{c}{2}\,\vf_\be(q)\vf_\ga(q)\,.
   }
 \end{array}
  \eq
\noindent 2. The change of variables (\ref{a251}) is a symplectic map, i.e. the Poisson brackets between $S_\al(p(x),q(x),c)$ computed via
the canonical bracket $\{p(x),q(y)\}=\delta(x-y)$ reproduce the Poisson-Lie structure (\ref{a232}), and the value of its Casimir function is defined by the constant $c$:
 \beq\label{a2512}
 \begin{array}{c}
   \displaystyle{
   \frac{1}{2}\sum\limits_{\al=1}^3 S^2_{\al}(p,q,c)=\frac{c^2}{4}\,,
   }
 \end{array}
  \eq
that is the eigenvalues of the matrix $S(p,q,c)$ are equal to $\pm c/2$:
 \beq\label{a2513}
 \begin{array}{c}
   \displaystyle{
   \lambda=\frac{c}{2}\,.
   }
 \end{array}
  \eq
\noindent 3. The Hamiltonians $\mH^{\rm CM}$ (\ref{a220}) and
$\mH^{\rm LL}$ (\ref{a2327}) coincide on (\ref{a251}). The equations of motion (\ref{a233}) and (\ref{a221})
are also related by the change of variables (\ref{a251}).

\noindent 4.  In the classical mechanics limit $k=0$ (and $\ti\nu=\nu=c$) the gauge equivalence
(\ref{a130})-(\ref{a134})
between the two-body Calogero-Moser model and the elliptic (XYZ) Euler top is reproduced.
\end{predl}

The proof of the above statements is straightforward though the calculations are cumbersome.
To verify the change of variables one should substitute the matrix (\ref{a248}) into (\ref{a250}), then use (\ref{A13})-(\ref{A16}) to
exclude the theta functions with doubled $\tau$, and apply the Riemann identities (\ref{A17})-(\ref{A24}) to
transform the obtained expressions to the form of the Landau-Lifshitz Lax operator (\ref{a1196}) with the
functions (\ref{A281}). For example, in this way one finds expressions
  \beq\label{B5}
  \begin{array}{c}
    \displaystyle{
  \det \rho{\ti g}(z)=\ti\nu(c+k q_{x})\vartheta(z) \vartheta(q)\,,
  }
  \end{array}
   \eq
   %
   %and
   %
  \beq\label{B8}
  \begin{array}{c}
    \displaystyle{
  {\ti g}_{11}(z) {\ti g}_{22}(z)+{\ti g}_{21}(z) {\ti g}_{12}(z)=
  \rho^{-2}\ti\nu(c+k q_{x})\theta_{10}(z) \theta_{10}(q)\,,
  }
  \end{array}
   \eq
and so on, which are necessary for computing ${\ti g}U^{CM}{\ti g}^{-1}$.
 Verification of other statements is partly simplified by using the compact form (\ref{a2511}).

%\subsection*{Real case}

%%%%%%%%%%%%%%%%%%%%%%%%%%%%%%%%%%%%%%%%%%%%%%%%%%%%%%%%%%%%%%%%%%%%%%%%%%%%%%%%%%%%%%%%%%%%%%%%%%%%%%
%%%%%%%%%%%%%%%%%%%%%%%%%%%%%%%%%%%%%%%%%%%%%%%%%%%%%%%%%%%%%%%%%%%%%%%%%%%%%%%%%%%%%%%%%%%%%%%%%%%%%%

\section{Trigonometric models}\label{sect3}
\setcounter{equation}{0}

In this Section we describe similar relations between the 2-body trigonometric 1+1 Calogero-Moser field theory and
the corresponding Landau-Lifshitz equation. On the side of the top or Landau-Lifshitz equation we deal
with the model related to 7-vertex $R$-matrix \cite{Chered}. The (deformed) Landau-Lifshitz model for
this $R$-matrix was described in \cite{KY}. From the very beginning we consider the field case, and the mechanics
 appears in $k=0$ limit. It can be found in \cite{KrZ}.

\paragraph{Trigonometric deformed (7-vertex XXZ) Landau-Lifshitz model.} Instead of the Pauli matrices basis here we use the standard one, so that the fields $S_{ij}(x)$ are components of the matrix
 \beq\label{a187}
 \begin{array}{c}
   \displaystyle{
 S(x)=\left(
\begin{array}{cc}
 S_{11}(x) & S_{12}(x) \\
 S_{21}(x) & -S_{11}(x) \\
\end{array}
\right)\,.
 }
 \end{array}
  \eq
The Poisson brackets are of the form:
 \beq\label{a191}
 \begin{array}{c}
   \displaystyle{
 \left\{S_{i j}(x), S_{k l}(y)\right\}=\left(S_{i l}(x) \delta_{k j}-S_{k j}(x) \delta_{i l}\right) \delta(x-y)\,.
 }
 \end{array}
  \eq
  The inverse inertia tensor responsible for anisotropy in the magnet takes the form
 \beq\label{a80}
 \begin{array}{c}
   \displaystyle{
 J(S)=\left(\begin{array}{cc}{0} & {-\frac{1}{2}\,S_{12}} \\ {-4 S_{12}-\frac{1}{2}\,S_{21}} & {0}\end{array}\right)\,.
}
 \end{array}
  \eq
  The Landau-Lifshitz equations
 \beq\label{a192}
 \begin{array}{c}
   \displaystyle{
 \partial_{t} S=[J(S), S]-\alpha_0\left[S, S_{x x}\right]\,,\qquad \al_0=k^2/8\lambda^2
 }
 \end{array}
  \eq
  are generated by the Hamiltonian
  \beq\label{a190}
  \begin{array}{c}
   \displaystyle{
  \mH^{\mathrm{LL}}=\frac{1}{2}
   \oint\mathrm{d} \mathrm{x} \Big(\tr(SJ(S))-\alpha_0 \tr( S_{x}^{2} ) \Big)=
  }
    \\ \ \\
        \displaystyle{
 =\frac{1}{2} \oint\mathrm{d} \mathrm{x} \Big(-4 S_{12}^2-S_{12} S_{21}-2\alpha_0  \left(\p_xS_{11}\p_xS_{11}+ \p_xS_{12}\p_xS_{21}\right) \Big)\,.
  }
  \end{array}
   \eq
   The zero curvature equation (\ref{a11}) is equivalent to (\ref{a192}) for the $U-V$ pair
 \beq\label{a184}
 \begin{array}{c}
   \displaystyle{
U^{LL}(z)=
\left(
\begin{array}{cc}
    \displaystyle{ S_{11} \coth  (z)} &    \displaystyle{\frac{S_{12}}{\sinh (z)} }
    \\
    \displaystyle{\frac{S_{21}}{\sinh (z)} -4  S_{12} \sinh  (z) } &    \displaystyle{ -S_{11} \coth  (z) }
\end{array}
\right)
 }
 \end{array}
  \eq
  and
 \beq\label{a1841}
 \begin{array}{c}
   \displaystyle{
    V^{\rm LL}(z)=\frac{1}{2}\,\Big( \coth(z)U^{\rm LL}(z,S)+2M^{\rm top}(z,S)-U^{\rm LL}(z,v) \Big)\,,
    }
 \end{array}
   \eq
where the matrix $v$ is defined as in (\ref{a1198}) and
 \beq\label{a1842}
 \begin{array}{c}
   \displaystyle{
    M^{\rm top}(z,S)=-\frac{1}{2}\mat{S_{11}}{0}{-8S_{12}\cosh(z)}{-S_{11}}
     }
 \end{array}
   \eq
is the accompany $M$-matrix entering the Lax equations in the classical mechanics.

\paragraph{1+1 two-body trigonometric Calogero-Moser field theory.}
The Hamiltonian is as follows:
 \beq\label{a179}
 \begin{array}{c}
   \displaystyle{
 \mH^{\rm CM}=\frac{1}{2}\oint {\rm dx}\left( {{p}^{2}}\left( 1-\frac{{{k}^{2}}q_{x}^{2}}{{{c}^{2}}} \right)+\frac{(3{{k}^{2}}q_{x}^{2}-{{c}^{2}})}{4}\coth ^2(q)-\frac{ k^2 q_{x}^{2}}{2}-\frac{{{k}^{4}}{{q}^{2}_{xx}}}{4{{{\tilde\nu} }^{2}}} \right)\,,
 }
 \end{array}
  \eq
where ${\tilde\nu}^{2}=c^2-k^2 q_{x}^{2}$ as in (\ref{a2216}). Equations of motion take the form:
 \beq\label{a181}
 \begin{array}{c}
   \displaystyle{
 \p_t q={p}\left( 1-\frac{{{k}^{2}}q_{x}^{2}}{{{c}^{2}}} \right)\,,
 }
 \end{array}
  \eq
  \beq\label{a182}
  \begin{array}{c}
    \displaystyle{
  \p_t p=-\frac{ k^2}{ c^2} \partial_{x}\left(p^{2} q_{x}\right)+\frac{(3{{k}^{2}}q_{x}^{2}-{{c}^{2}})\coth(q)}{4\sinh^{2}(q)}+\frac{3 k^2}{4} \partial_{x}\left( q_{x} \coth ^2(q) \right)-
  }
  \\ \ \\
   \displaystyle{
  -\frac{ k^2 q_{xx}}{2}+\frac{k^4}{4} \partial_{x}\left(\frac{q_{x x x} {\tilde\nu}-{\tilde\nu}_{x} q_{x x}}{{\tilde\nu}^{3}}\right)\,.
  }
  \end{array}
   \eq
   They are represented in the Zakharov-Shabat form (\ref{a11}) with
 \beq\label{a175}
 \begin{array}{c}
   \displaystyle{
 U^{\rm CM}(z)=\left(
\begin{array}{cc}
   \displaystyle{p -\frac{{k {q}_{x}}}{2}\coth (z) }
   &
     \displaystyle{ \frac{{\tilde\nu}}{2}  \coth (q)-\frac{{\tilde\nu}}{2}  \coth (z) }
  \\ \ \\
  \displaystyle{ -\frac{{\tilde\nu}}{2}  \coth (q)-\frac{{\tilde\nu}}{2}  \coth (z) }
   &
     \displaystyle{ -p+\frac{{k {q}_{x}}}{2}\coth (z) }
\end{array}
\right)
 }
 \end{array}
  \eq
and
 \beq\label{a1750}
 \begin{array}{c}
   \displaystyle{
  V^{\rm CM}(z)=\mat{V_{11}^{\rm CM}(z)}{V_{12}^{\rm CM}(z)}{V_{21}^{\rm CM}(z)}{-V_{11}^{\rm CM}(z)}
   }
 \end{array}
  \eq
with the following matrix elements:
 \beq\label{a1751}
 \begin{array}{c}
   \displaystyle{
   V_{11}^{\rm CM}(z)=\frac14\left(2q_t\coth(z)+\frac{4k}{c^2}p^2q_x-3kq_x\coth^2(q)
   +2kq_x-\right.
   }
   \\ \ \\
    \displaystyle{
  \left. -k^3\Big(\frac{q_{x x x} {\tilde\nu}-{\tilde\nu}_{x} q_{x x}}{{\tilde\nu}^{3}}\Big)+\frac{kq_x}{\sinh^2(q)}-\frac{kq_x}{\sinh^2(z)}\right)\,,
   }
 \end{array}
  \eq
 \beq\label{a1753}
 \begin{array}{c}
    \displaystyle{
 V_{12}^{\rm CM}(z)= -\frac{\ti\nu}{4\sinh^2(q)}-\Big(\frac{\ti\nu}{4}\coth(z)+\frac{kpq_x\ti\nu}{2c^2}
 +\frac{k^2q_{xx}}{4\ti\nu}\Big)(\coth(z)-\coth(q))\,,
 }
 \end{array}
  \eq
 \beq\label{a1754}
 \begin{array}{c}
    \displaystyle{
 V_{21}^{\rm CM}(z)= -\frac{\ti\nu}{4\sinh^2(q)}-\Big(\frac{\ti\nu}{4}\coth(z)+\frac{kpq_x\ti\nu}{2c^2}
 -\frac{k^2q_{xx}}{4\ti\nu}\Big)(\coth(z)+\coth(q))\,.
 }
 \end{array}
  \eq

\paragraph{Gauge equivalence and change of variables.}
First, let us obtain expression for the matrix of the gauge transformation $\ti g(z)$.
 Similarly to elliptic case we represent it in the form (\ref{a2010}) $\ti g(z)=\Xi(z){\ti D}^{-1}$.
In the elliptic case the $\Xi(z)$ matrix was the intertwining matrix for the 8-vertex (discrete XYZ) model.
In the trigonometric case we deal with the intertwining matrix for the 7-vertex (deformed discrete XXZ) model.
It is of the form \cite{AHZ,KrZ}:
 \beq\label{a205}
 \begin{array}{c}
   \displaystyle{
\Xi(z)=\frac{1}{2}\left(
\begin{array}{cc}
 1 & 1 \\
 2 \cosh  ( z+q  ) & 2 \cosh  (z-q ) \\
\end{array}
\right)\,.
}
 \end{array}
  \eq
 The rest of derivation is just the same as in the elliptic case. Since
 the residue of $U^{\rm CM}(z)$ (\ref{a175}) is equal to the one (\ref{a2012}) we have
 \beq\label{a207}
 \begin{array}{c}
   \displaystyle{
{\ti D}=\rho\left( \begin{array}{cc}
    \frac{1}{{\tilde\nu} }  & 0 \\
0 &     -\frac{1}{k{{q}_{x}}+c}
\end{array} \right)\,,
 }
 \end{array}
  \eq
where the normalization factor $\rho$ comes from the condition $\det\ti g=1$. This yields
% in order to have
% $U^{\rm CM}(z)\,,U^{\rm LL}(z)\in {\rm sl}(2,\mC)$:
%
  %
 \beq\label{a208}
 \begin{array}{c}
   \displaystyle{
 {\ti g}(z)=\frac{1}{\rho}\left(
\begin{array}{cc}
 \displaystyle{ \frac{{\tilde\nu} }{2} }& \displaystyle{ -\frac{1}{2} \left(c+k q_x\right) }
  \\ \ \\
 \displaystyle{\cosh (z+q) {\tilde\nu} } &  \displaystyle{- \cosh (z-q) \left(c+k q_x\right) }
\end{array}
\right)\,,
 }
 \end{array}
  \eq
 \beq\label{a2081}
 \begin{array}{c}
   \displaystyle{
\rho=\sqrt{ \ti\nu(c+k q_{x})\sinh(z) \sinh(q) }\,.
 }
 \end{array}
  \eq
The main statement in this Section is absolutely parallel to the one for elliptic case, see the Proposition \ref{predl1}.
The change of variables is given by
 \beq\label{a212}
 \begin{array}{c}
   \displaystyle{
S_{11}=-\Big(p-\frac{c}{2}\,\frac{ k^2 q_{xx}}{c^2-k^2 q_{x}^2} \Big)\coth (q)-\frac{c}{2}\,\frac1{\sinh^2(q)}\,,
 }
 \end{array}
  \eq
 \beq\label{a211}
 \begin{array}{c}
   \displaystyle{
S_{12}=\Big(p-\frac{c}{2}\,\frac{ k^2 q_{xx}}{c^2-k^2 q_{x}^2} \Big)\frac{1}{2\sinh(q)}+\frac{c}{4}\,\frac{\cosh(q)}{\sinh^2(q)}\,,
 }
 \end{array}
  \eq
 \beq\label{a213}
 \begin{array}{c}
   \displaystyle{
S_{21}=-\Big(p-\frac{c}{2}\,\frac{ k^2 q_{xx}}{c^2-k^2 q_{x}^2} \Big)\frac{2\cosh^2(q)}{\sinh(q)}
+2c\,\cosh(q)-c\frac{\cosh^3(q)}{\sinh^2(q)}\,,
 }
 \end{array}
  \eq
or similarly to (\ref{a2511}) we have a compact form:
 \beq\label{a2124}
 \begin{array}{c}
   \displaystyle{
S_{ij}=\Big(p-\frac{c}{2}\,\frac{ k^2 q_{xx}}{c^2-k^2 q_{x}^2}-\frac{c}{2}\,\p_q \Big)\vf_{ij}(q)\,,
 }
 \end{array}
  \eq
  where
 \beq\label{a2115}
 \begin{array}{c}
   \displaystyle{
 \vf_{11}(q)=-\coth (q)\,,\quad \vf_{12}(q)=\frac{1}{2\sinh(q)}\,,\quad \vf_{21}(q)=-\frac{2\cosh^2(q)}{\sinh(q)}\,.
 }
 \end{array}
  \eq
The Poisson brackets computed through the canonical bracket (\ref{a222}) has the form (\ref{a191}):
 \beq\label{a214}
 \begin{array}{l}
   \displaystyle{
\left\{S_{1 1}(x), S_{1 2}(y)\right\}=S_{1 2}(x) \delta(x-y)\,,
 }
\\ \ \\
   \displaystyle{
\left\{S_{1 1}(x), S_{2 1}(y)\right\}=-S_{2 1}(x) \delta(x-y)\,,
 }
\\ \ \\
   \displaystyle{
\left\{S_{1 2}(x), S_{2 1}(y)\right\}=2S_{1 1}(x) \delta(x-y)\,,
 }
 \end{array}
  \eq
and the eigenvalues of the matrix $S$ are the constants $\pm\lambda=\pm c/2$.

%%%%%%%%%%%%%%%%%%%%%%%%%%%%%%%%%%%%%%%%%%%%%%%%%%%%%%%%%%%%%%%%%%%%%%%%%%%%%%%%%%%%%%%%%%%%%%%%%%%%%%
%%%%%%%%%%%%%%%%%%%%%%%%%%%%%%%%%%%%%%%%%%%%%%%%%%%%%%%%%%%%%%%%%%%%%%%%%%%%%%%%%%%%%%%%%%%%%%%%%%%%%%

\section{Rational models}\label{sect4}
\setcounter{equation}{0}
Here we describe the gauge equivalence between the 2-body rational 1+1 Calogero-Moser filed theory
and the Landau-Lifshitz equation related to 11-vertex model \cite{LOZ11} based on the deformed XXX $R$-matrix from \cite{Chered}.

\paragraph{Rational deformed (11-vertex XXX) Landau-Lifshitz model.}
We keep notations (\ref{a187}) and the form of the Poisson brackets (\ref{a191})
from the previous Section. In the rational case we have the following anisotropy matrix
$J(S)$:
 \beq\label{a1451}
 \begin{array}{c}
   \displaystyle{
J(S)=-\mat{S_{12}}{0}{2S_{11}}{-S_{12}}\,.
 }
 \end{array}
  \eq
The Landau-Lifshitz equation (\ref{a192}) with $J(S)$ (\ref{a1451}) is generated by the Hamiltonian:
  \beq\label{a148}
  \begin{array}{c}
  \displaystyle{
  \mH^{\mathrm{LL}}=\frac{1}{2}
   \oint\mathrm{d} \mathrm{x} \Big(\tr(SJ(S))-\alpha_0 \tr( S_{x}^{2} ) \Big)=
  }
    \\ \ \\
    \displaystyle{
  =\frac{1}{2} \oint\mathrm{d} \mathrm{x} \Big(-4 S_{11} S_{12}-2\alpha_0  \left( (S_{11})_{x}^2+ (S_{12})_{x} (S_{21})_{x}\right) \Big)\,.
  }
  \end{array}
   \eq
The U-V pair has the form \cite{LOZ11}:
 \beq\label{a142}
 \begin{array}{c}
   \displaystyle{
U^{\rm LL}(z)=\left(
\begin{array}{cc}
\displaystyle{ \frac{S_{11}}{z}-z S_{12} } & \displaystyle{ \frac{S_{12}}{z} }
 \\ \ \\
 \displaystyle{\frac{S_{21}}{z} -2z S_{11} -z^3 S_{12} } & \displaystyle{ -\frac{S_{11}}{z}+z S_{12} }
\end{array}
\right)\,,
 }
 \end{array}
  \eq
 \beq\label{a1426}
 \begin{array}{c}
   \displaystyle{
    V^{\rm LL}(z)=\frac{1}{2}\,\Big( \frac{1}{z}\,U^{\rm LL}(z,S)+2M^{\rm top}(z,S)-U^{\rm LL}(z,v) \Big)\,,
    }
 \end{array}
   \eq
where
 \beq\label{a14518}
 \begin{array}{c}
   \displaystyle{
M^{\rm top}(z,S)=\mat{S_{12}}{0}{2S_{11}+2z^2S_{12}}{-S_{12}}\,.
 }
 \end{array}
  \eq

\paragraph{1+1 two-body trigonometric Calogero-Moser field theory.}
 The canonical Poisson bracket $\left\{p(x), q(y)\right\}= \delta(x-y)$
and the Hamiltonian
 \beq\label{a138}
 \begin{array}{c}
   \displaystyle{
 \mH^{\rm CM}=\frac12\oint {\rm d x}\left({{p}^{2}}\left( 1-\frac{{{k}^{2}}q_{x}^{2}}{{{c}^{2}}} \right)+\frac{3{{k}^{2}}q_{x}^{2}-{{c}^{2}}}{4{{q}^{2}}}-\frac{{{k}^{4}}{{q}^{2}_{xx}}}{4{{{\tilde\nu} }^{2}}}\right)\,,
 }
 \end{array}
  \eq
  where $\ti\nu=\sqrt{c^2-k^2q_x^2}$,
provides equations of motion
  \beq\label{a140}
  \begin{array}{c}
    \displaystyle{
  \dot{q}={p}\left( 1-\frac{{{k}^{2}}q_{x}^{2}}{{{c}^{2}}} \right)\,,
  }
  \\ \ \\
   \displaystyle{
  \dot{p}=-\frac{ k^2}{ c^2} \partial_{x}\left(p^{2} q_{x}\right)+\frac{3{{k}^{2}}q_{x}^{2}-{{c}^{2}}}{4{{q}^{3}}}+\frac{3 k^2}{4} \partial_{x}\left( \frac{q_{x}}{ q^2} \right)+\frac{k^4}{4} \partial_{x}\left(\frac{q_{x x x} {\tilde\nu}-{\tilde\nu}_{x} q_{x x}}{{\tilde\nu}^{3}}\right)\,.
  }
  \end{array}
   \eq
These equations are represented in the Zakharov-Shabat form with the following $U-V$ pair:
 \beq\label{a136}
 \begin{array}{c}
   \displaystyle{
U^{\rm CM}(z)=\left( \begin{array}{cc}
\displaystyle{p -\frac{{k{q}_{x}}}{2z} } &
\displaystyle{ -{\tilde\nu} \left( \frac{1}{2z}-\frac{1}{2q} \right) }
\\ \ \\
\displaystyle{ -{\tilde\nu} \left( \frac{1}{2z}+\frac{1}{2q} \right) }&
\displaystyle{ -p +\frac{k{{q}_{x}}}{2z} }
\end{array} \right)
 }
 \end{array}
  \eq
  and $V^{\rm CM}(z)$ with the elements
 \beq\label{a1367}
\left\{\begin{array}{l}
   \displaystyle{
V_{11}^{\rm CM}(z)=\frac{q_{t}}{2 z} -\left(-\frac{ k}{ c^2} p^{2} q_{x}+\frac{3 k q_{x}}{4 q^{2}} +\frac{k^3}{4} \left(\frac{q_{x x x} \nu-\nu_{x} q_{x x}}{\nu^{3}}\right)\right)
+\frac{k q_{x}}{4}\left(\frac{1}{q^{2}}-\frac{1}{z^{2}}\right)\,,
}
 \\ \ \\
    \displaystyle{
V_{12}^{\rm CM}(z)=-\frac{\nu}{4 q^{2}} +\left(-\frac{\nu}{4 z} -\frac{k p q_{x} \nu}{2 c^{2} }-\frac{k^{2} q_{x x}}{4  \nu}\right) \left(\frac{1}{z}-\frac{1}{q}\right)\,,
}
 \\ \ \\
    \displaystyle{
V_{21}^{\rm CM}(z)=-\frac{\nu}{4 q^{2}}+\left(-\frac{\nu}{4 z} -\frac{k p q_{x} \nu}{2 c^{2} }+\frac{k^{2} q_{x x}}{4  \nu}\right)\left(\frac{1}{z}+\frac{1}{q}\right)\,.
}
\end{array}\right.
  \eq

\paragraph{Gauge equivalence and change of variables.}
The matrix of the gauge transformation is constructed similarly to the previous Sections.
Put $\ti g(z)=\Xi(z){\ti D}^{-1}$. Here $\Xi(z)$ is the intertwining matrix
for the 11-vertex model \cite{AASZ,LOZ2}:
 \beq\label{a2057}
 \begin{array}{c}
   \displaystyle{
\Xi(z)=\frac{1}{2}\left(
\begin{array}{cc}
 1 & 1 \\
 (z+q  )^2 & (z-q )^2 \\
\end{array}
\right)\,.
}
 \end{array}
  \eq
 Again, the rest of derivation is just the same as in the elliptic case. Since
 the residue of $U^{\rm CM}(z)$ (\ref{a136}) is equal to the one (\ref{a2012}) we have
 \beq\label{a2077}
 \begin{array}{c}
   \displaystyle{
{\ti D}=\rho\left( \begin{array}{cc}
    \frac{1}{{\tilde\nu} }  & 0 \\
0 &     -\frac{1}{k{{q}_{x}}+c}
\end{array} \right)\,,
 }
 \end{array}
  \eq
where the normalization factor $\rho$ is evaluated from the requirement for $\ti g$
% in order to have
% $U^{\rm CM}(z)\,,U^{\rm LL}(z)\in {\rm sl}(2,\mC)$:
%
  %
 \beq\label{a2087}
 \begin{array}{c}
   \displaystyle{
 {\ti g}(z)=\frac{1}{\rho}\left(
\begin{array}{cc}
 \displaystyle{ \frac{{\tilde\nu} }{2} }& \displaystyle{ -\frac{1}{2} \left(c+k q_x\right) }
  \\ \ \\
 \displaystyle{(z+q  )^2 {\tilde\nu} } &  \displaystyle{- (z-q  )^2 \left(c+k q_x\right) }
\end{array}
\right)\,,
 }
 \end{array}
  \eq
  to be the element of ${\rm SL}(2,\mC)$, that is
 \beq\label{a20817}
 \begin{array}{c}
   \displaystyle{
\rho=\sqrt{ \ti\nu(c+k q_{x})zq }\,.
 }
 \end{array}
  \eq
Then the changes of the field variables follows from (\ref{a250}):
 \beq\label{a170}
 \begin{array}{l}
   \displaystyle{
S_{11}=-\frac{p q}{2}+\frac{c}{4}+\frac{c k^2 q q_{xx}}{4 \left(c^2-k^2 q_x^2\right)}\,,
 }
\\ \ \\
   \displaystyle{
{{S}_{12}}=\frac{p}{2 q}+\frac{c}{4 q^2}-\frac{c k^2 q_{xx}}{4 q \left(c^2-k^2 q_x^2\right)}\,,
 }
\\ \ \\
   \displaystyle{
S_{21}=-\frac{p q^3}{2}+\frac{3 c q^2}{4}+\frac{c k^2 q^3 q_{xx}}{4 \left(c^2-k^2 q_x^2\right)}\,.
 }
 \end{array}
  \eq
They are rewritten in the compact form:
 \beq\label{a1701}
 \begin{array}{c}
   \displaystyle{
S_{ij}=\Big(p-\frac{c}{2}\,\frac{ k^2 q_{xx}}{c^2-k^2 q_{x}^2}-\frac{c}{2}\,\p_q \Big)\vf_{ij}(q)\,,
 }
 \end{array}
  \eq
  where
 \beq\label{a1702}
 \begin{array}{c}
   \displaystyle{
 \vf_{11}(q)=-\frac{q}{2}\,,\quad \vf_{12}(q)=\frac{1}{2q}\,,\quad \vf_{21}(q)=-\frac{q^3}{2}\,.
 }
 \end{array}
  \eq
  %

%%%%%%%%%%%%%%%%%%%%%%%%%%%%%%%%%%%%%%%%%%%%%%%%%%%%%%%%%%%%%%%%%%%%%%%%%%%%%%%%%%%%%%%%%%%%%%%%%%%%%%
%%%%%%%%%%%%%%%%%%%%%%%%%%%%%%%%%%%%%%%%%%%%%%%%%%%%%%%%%%%%%%%%%%%%%%%%%%%%%%%%%%%%%%%%%%%%%%%%%%%%%%

%\section{Relation to Nonlinear Schrödinger equation}
%\setcounter{equation}{0}

%%%%%%%%%%%%%%%%%%%%%%%%%%%%%%%%%%%%%%%%%%%%%%%%%%%%%%%%%%%%%%%%%%%%%%%%%%%%%%%%%%%%%%%%%%%%%%%%%%%%%%
%%%%%%%%%%%%%%%%%%%%%%%%%%%%%%%%%%%%%%%%%%%%%%%%%%%%%%%%%%%%%%%%%%%%%%%%%%%%%%%%%%%%%%%%%%%%%%%%%%%%%%

\section{Conclusion}
\setcounter{equation}{0}
Based on results of \cite{Kr2,LOZ} we explicitly constructed
the gauge transformation relating the field generalization of 2-body Calogero-Moser-Sutherland models and the Landau-Lifshitz models
in the elliptic, trigonometric and rational cases.
The answer for the gauge transformation partly  comes from mechanics, where it is given by the IRF-Vertex transformation for
the 8-vertex, 7-vertex and 11-vertex models respectively. Their field generalizations ${\ti g}(z)$ are derived from conditions
for modifications of underlying bundles with the loop (structure) group over elliptic curve and its trigonometric (nodal) and rational (cuspidal) degenerations. The answers are given by (\ref{a248}), (\ref{a208}) and (\ref{a2087}).

By applying the gauge transformation we obtain explicit changes of variables between the Calogero-Moser
 field theories and the Landau-Lifshitz equations. Results can be represented in the following compact form:
 \beq\label{a25111}
 \begin{array}{c}
   \displaystyle{
   S_{\al}(p,q,c) =  \Big(p-\frac{c}{2}\,\frac{k^2q_{xx}}{c^2-k^2q_x^2}-\frac{c}{2}\,\p_q\Big)h_\al(q)\,,
   }
 \end{array}
  \eq
where for the elliptic case $\al=1,2,3$ -- the indices  of the Pauli matrices basis and $h_\al(q)=c_\al(\tau)\vf_\al(q)$
(\ref{a2511}). In the trigonometric and rational cases $\al=11,12,21$ - the indices  of the standard basis, and
the sets of functions $h_\al(q)$ are given by (\ref{a2115}) and (\ref{a1702}) respectively.

Thus, we may formulate a simple receipt
for the field generalizations of the formulae in mechanics. Suppose we have a set changes of variables
$S^{\rm mech}_\al(p,q,\nu)$ between the 2-body Calogero-Moser model and the Euler top in the classical mechanics.
Then its field generalization
is of the form:
 \beq\label{a7007}
 \begin{array}{c}
   \displaystyle{
 S^{\rm field}_\al(p(x),q(x),c)=S^{\rm mech}_\al(\check{p}(x),q(x),c)\,,\quad
 \check{p}(x)=p(x)-\frac{c}{2}\,\frac{ k^2 q_{xx}}{c^2-k^2 q_{x}^2}\,.
 }
 \end{array}
  \eq
The results were obtained using the approach of \cite{LOZ}, which works in the ${\rm GL}(N,\mC)$ case \cite{LOSZ0} and can be
naturally extended to
a complex simple Lie group \cite{LOSZ}. We will study the field generalizations for these cases in our future articles.

%%%%%%%%%%%%%%%%%%%%%%%%%%%%%%%%%%%%%%%%%%%%%%%%%%%%%%%%%%%%%%%%%%%%%%%%%%%%%%%%%%%%%%%%%%%%%%%%%%%%%%
%%%%%%%%%%%%%%%%%%%%%%%%%%%%%%%%%%%%%%%%%%%%%%%%%%%%%%%%%%%%%%%%%%%%%%%%%%%%%%%%%%%%%%%%%%%%%%%%%%%%%%

\section{Appendix}
\def\theequation{A.\arabic{equation}}
\setcounter{equation}{0}

Here we give the definitions and properties of elliptic functions used in the paper.
Using the Riemann notations let us begin with the theta functions definition:
%
%Theta functions with characteristics.
%
%
  \beq\label{A11}
  \begin{array}{c}
    \displaystyle{
  \theta\left[\begin{array}{l}
a \\
b
\end{array}\right](z, \tau)=\sum_{j \in \mathbb{Z}} \mathbf{e}\left((j+a)^{2} \frac{\tau}{2}+(j+a)(z+b)\right)\,,
\quad {\rm Im}(\tau)>0\,.
  }
  \end{array}
   \eq
In this paper the characteristics $a,b$ are either $0$ or $1/2$. For shortness we denote $\theta\left[\begin{array}{l}
a / 2 \\
b / 2
\end{array}\right]=\theta_{a b}$. The odd theta function is also denoted as
$\vartheta(x)$, so that
  \beq\label{A12}
  \begin{array}{c}
    \displaystyle{
\theta_{11}(x, \tau)=\vartheta(x)=\theta\left[\begin{array}{c}
1 / 2 \\
1 / 2
\end{array}\right](x, \tau)\,.
}
  \end{array}
   \eq
Relation to the Jacobi notations is as follows:
  \beq\label{A1}
  \begin{array}{c}
    \displaystyle{
 \vth(z)=-\theta_1(z)\,,\quad
 \theta_{10}(z)=\theta_2(z)\,,\quad
 \theta_{00}(z)=\theta_3(z)\,,\quad
 \theta_{01}(z)=\theta_4(z)\,.
  }
  \end{array}
   \eq
   The Eisenstein functions:
  \beq\label{A2}
  \begin{array}{c}
    \displaystyle{
  E_{1}(z | \tau)=\partial_{z} \log \vartheta(z | \tau)\,,
  }
  \end{array}
   \eq
  \beq\label{A4}
  \begin{array}{c}
    \displaystyle{
  E_{2}(z | \tau)=-\partial_{z} E_{1}(z | \tau)=\partial_{z}^{2} \log \vartheta(z | \tau)=
  \wp(z)-\frac{1}{3}\frac{\vth'''(0)}{\vth'(0)}\,,
  }
  \end{array}
   \eq
 The Kronecker elliptic function and its derivative:
  \beq\label{A5}
  \begin{array}{c}
    \displaystyle{
  \phi(z,u)=\frac{ \vartheta'(0)\vartheta(z+u)}{\vartheta(z)\vartheta(u)}\,,
  }
  \end{array}
   \eq
  \beq\label{A61}
  \begin{array}{c}
    \displaystyle{
  f(z,u)=\p_u\phi(z,u)=\phi(z,u)(E_1(z+u)-E_1(u))\,.
  }
  \end{array}
   \eq
Addition formulae (or the Fay identity):
  \beq\label{A9}
  \begin{array}{c}
    \displaystyle{
  \phi(z,u) \partial_{v} \phi(z,v)-\phi(z,v) \partial_{u} \phi(z,u)=\left(E_{2}(u)-E_{2}(v)\right) \phi(z,u+v)\,,
  }
  \end{array}
   \eq
  \beq\label{A10}
  \begin{array}{c}
    \displaystyle{
  \phi (z,-q) \phi (z,q)=\wp(z)-\wp(q)=E_{2}(z)-E_{2}(q)\,.
  }
  \end{array}
   \eq
Addition formulae (Riemann identities for theta functions):
  \beq\label{A17}
  \begin{array}{c}
    \displaystyle{
  \vartheta(x+u) \theta_{00}(x-u) \theta_{01}(0) \theta_{10}(0)=\theta_{00}(x) \vartheta(x) \theta_{01}(u) \theta_{10}(u)+
\theta_{10}(x) \theta_{01}(x) \theta_{00}(u) \vartheta(u)\,,
  }
  \end{array}
   \eq
  \beq\label{A18}
  \begin{array}{c}
    \displaystyle{
  \vartheta(x-u) \theta_{00}(x+u) \theta_{01}(0) \theta_{10}(0)=\theta_{00}(x) \vartheta(x) \theta_{01}(u) \theta_{10}(u)-
 \theta_{10}(x) \theta_{01}(x) \theta_{00}(u) \vartheta(u)\,,
  }
  \end{array}
   \eq
  \beq\label{A19}
  \begin{array}{c}
    \displaystyle{
  \vartheta(x+u) \theta_{01}(x-u) \theta_{00}(0) \theta_{10}(0)=\theta_{00}(x) \theta_{10}(x) \theta_{01}(u) \vartheta(u)+
  \theta_{01}(x) \vartheta(x) \theta_{00}(u) \theta_{10}(u)\,,
  }
  \end{array}
   \eq
  \beq\label{A20}
  \begin{array}{c}
    \displaystyle{
  \vartheta(x-u) \theta_{01}(x+u) \theta_{00}(0) \theta_{10}(0)=-\theta_{00}(x) \theta_{10}(x) \theta_{01}(u) \vartheta(u)+
  \theta_{01}(x) \vartheta(x) \theta_{00}(u) \theta_{10}(u)\,,
  }
  \end{array}
   \eq
  \beq\label{A21}
  \begin{array}{c}
    \displaystyle{
  \vartheta(x+u) \theta_{10}(x-u) \theta_{00}(0) \theta_{01}(0)=\theta_{00}(x) \theta_{01}(x) \theta_{10}(u) \vartheta(u)+
  \theta_{01}(x) \vartheta(x) \theta_{00}(u) \theta_{10}(u)\,,
  }
  \end{array}
   \eq
  \beq\label{A22}
  \begin{array}{c}
    \displaystyle{
  \vartheta(x-u) \theta_{10}(x+u) \theta_{00}(0) \theta_{01}(0)=-\theta_{00}(x) \theta_{01}(x) \theta_{10}(u) \vartheta(u)+
  \theta_{10}(x) \vartheta(x) \theta_{00}(u) \theta_{01}(u)\,.
  }
  \end{array}
   \eq
 In particular cases one gets relations between the squares of theta functions:
  \beq\label{A23}
  \begin{array}{c}
    \displaystyle{
  \vartheta(z)^{2} \theta_{01}(0)^{2}=\theta_{00}(z)^{2} \theta_{10}(0)^{2}-\theta_{10}(z)^{2} \theta_{00}(0)^{2}\,,
  }
  \end{array}
   \eq
  \beq\label{A231}
  \begin{array}{c}
   \displaystyle{
  \theta_{10}(z)^{2} \theta_{01}(0)^{2}=\theta_{01}(z)^{2} \theta_{10}(0)^{2}-\vartheta(z)^{2} \theta_{00}(0)^{2}\,,
  }
  \end{array}
   \eq
  \beq\label{A232}
  \begin{array}{c}
  \displaystyle{
  \theta_{00}(z)^{2} \theta_{01}(0)^{2}=\theta_{01}(z)^{2} \theta_{00}(0)^{2}-\vartheta(z)^{2} \theta_{10}(0)^{2}\,,
  }
  \end{array}
   \eq
  \beq\label{A233}
  \begin{array}{c}
   \displaystyle{
  \theta_{01}(z)^{2} \theta_{01}(0)^{2}=\theta_{00}(z)^{2} \theta_{00}(0)^{2}-\theta_{10}(z)^{2} \theta_{10}(0)^{2}\,,
  }
  \end{array}
   \eq
  \beq\label{A25}
  \begin{array}{c}
    \displaystyle{
  \theta_{01}(x)^4+\theta_{10}(x)^4-\theta_{00}(x)^4=\theta_{11}(x)^4\,,
  }
  \end{array}
   \eq
  \beq\label{A24}
  \begin{array}{c}
    \displaystyle{
  \theta_{00}(0)^{4}=\theta_{10}(0)^{4}+\theta_{01}(0)^{4}\,.
  }
  \end{array}
   \eq
  Quasi-periodicity
Behavior on the lattice of periods $\mZ\oplus\tau\mZ$:
  \beq\label{A8}
  \begin{array}{c}
    \displaystyle{
  \vartheta(z+1)=-\vartheta(z), \quad \vartheta(z+\tau)=-q^{-\frac{1}{2}} e^{-2 \pi i z} \vartheta(z)\,,
  }
  \\ \ \\
   \displaystyle{
  E_{1}(z+1)=E_{1}(z), \quad E_{1}(z+\tau)=E_{1}(z)-2 \pi i\,,
  }
  \\ \ \\
   \displaystyle{
  E_{2}(z+1)=E_{2}(z), \quad E_{2}(z+\tau)=E_{2}(z)\,,
  }
  \\ \ \\
   \displaystyle{
  \phi(u+1, z)=\phi(u, z), \quad \phi(u+\tau, z)=e^{-2 \pi i z} \phi(u, z)\,,
  }
  \end{array}
   \eq
   \beq\label{A81}
 \begin{array}{c}
  \displaystyle{
\theta{\left[\begin{array}{c}
a\\
b
\end{array}
\right]}(z+1|\,\tau )=\exp(2\pi\imath a)\,
\theta{\left[\begin{array}{c}
a\\
b
\end{array}
\right]}(z |\,\tau )\,,
 }
 \end{array}
 \eq
 \beq\label{A82}
 \begin{array}{c}
  \displaystyle{
\theta{\left[\begin{array}{c}
a\\
b
\end{array}
\right]}(z+a'\tau|\,\tau ) =\exp\left(-2\pi\imath {a'}^2\frac\tau2
-2\pi\imath a'(z+b)\right) \theta{\left[\begin{array}{c}
a+a'\\
b
\end{array}
\right]}(z|\,\tau )\,.
 }
 \end{array}
 \eq

Theta functions with double modular parameter:
  \beq\label{A13}
  \begin{array}{c}
    \displaystyle{
  2 \vartheta(x, 2 \tau) \theta_{01}(y, 2 \tau)= \vartheta\left(\frac{x+y}{2}, \tau\right) \theta_{10}\left(\frac{x-y}{2}, \tau\right)
  +\theta_{10}\left(\frac{x+y}{2}, \tau\right) \vartheta\left(\frac{x-y}{2}, \tau\right)\,,
  }
  \end{array}
   \eq
  \beq\label{A14}
  \begin{array}{c}
    \displaystyle{
  2 \theta_{00}(x, 2 \tau) \theta_{10}(y, 2 \tau) =\vartheta\left(\frac{x+y}{2}, \tau\right) \vartheta\left(\frac{x-y}{2}, \tau\right)
  +\theta_{10}\left(\frac{x+y}{2}, \tau\right) \theta_{10}\left(\frac{x-y}{2}, \tau\right)\,,
  }
  \end{array}
   \eq
  \beq\label{A15}
  \begin{array}{c}
    \displaystyle{
  2 \theta_{00}(x, 2 \tau) \theta_{00}(y, 2 \tau) =\theta_{00}\left(\frac{x+y}{2}, \tau\right) \theta_{00}\left(\frac{x-y}{2}, \tau\right)
  +\theta_{01}\left(\frac{x+y}{2}, \tau\right) \theta_{01}\left(\frac{x-y}{2}, \tau\right)\,,
  }
  \end{array}
   \eq
  \beq\label{A16}
  \begin{array}{c}
    \displaystyle{
  2 \theta_{10}(x, 2 \tau) \theta_{10}(y, 2 \tau)=\theta_{00}\left(\frac{x+y}{2}, \tau\right) \theta_{00}\left(\frac{x-y}{2}, \tau\right)
  -\theta_{01}\left(\frac{x+y}{2}, \tau\right) \theta_{01}\left(\frac{x-y}{2}, \tau\right)\,.
  }
  \end{array}
   \eq
The quasiperiodic functions:
% $\varphi_{\alpha}(z)$ are numerated as follows:
%
  \beq\label{A26}
  \begin{array}{c}
    \displaystyle{
  \varphi_{\alpha}(z)=\exp(2\pi\imath\p_\tau\om_\al)\phi(z,\om_\al)\,,\quad \al=1,2,3\,,
  }
  \end{array}
   \eq
 where $\om_\al$ are half-periods of the elliptic curve (in the fundamental parallelogram)
  \beq\label{A261}
  \begin{array}{c}
    \displaystyle{
  \om_1=\frac{\tau}2\,,\quad \om_2=\frac{1+\tau}2\,,\quad \om_3=\frac12\,.
  }
  \end{array}
   \eq
More precisely,
  \beq\label{A281}
  \begin{array}{c}
    \displaystyle{
      \varphi_{1}(z)=\frac{\vartheta^{\prime}(0) \theta_{01}(z)}{\vartheta(z) \theta_{01}(0)}\,, \quad
      \varphi_{2}(z)=\frac{\vartheta^{\prime}(0) \theta_{00}(z)}{\vartheta(z) \theta_{00}(0)}\,,\quad
  \varphi_{3}(z)=\frac{\vartheta^{\prime}(0) \theta_{10}(z)}{\vartheta(z) \theta_{10}(0)}\,,
  }
  \end{array}
   \eq
 or
  \beq\label{A28}
  \begin{array}{c}
    \displaystyle{
  \varphi_{1}(z)=e^{\pi\imath z}\phi\Big(z,\frac{\tau}{2}\Big)\,, \quad
      \varphi_{2}(z)=e^{\pi\imath z}\phi\Big(z,\frac{1+\tau}{2}\Big)\,,\quad
  \varphi_{3}(z)=\phi\Big(z,\frac{1}{2}\Big)\,.
  }
  \end{array}
   \eq
   From (\ref{A9})-(\ref{A10}) we have (for distinct $\al,\be,\ga$)
  \beq\label{A31}
  \begin{array}{c}
    \displaystyle{
  \varphi_{\beta}(z) f_{\gamma}(z)-\varphi_{\gamma}(z) f_{\beta}(z) = \varphi_{\al}(z)\left(\wp\left(\omega_{\beta}\right)-\wp\left(\omega_{\gamma}\right)\right)\,,
  }
  \end{array}
   \eq
  \beq\label{A27}
  \begin{array}{c}
    \displaystyle{
  \varphi_{\alpha}(z)^2=\wp(z)-\wp\left(\omega_{\alpha}\right)\,.
  }
  \end{array}
   \eq
   where using the definition (\ref{A61}) we introduced
  \beq\label{A282}
  \begin{array}{c}
    \displaystyle{
  f_{1}(z)=e^{\pi\imath z}f\Big(z,\frac{\tau}{2}\Big)\,, \quad
      f_{2}(z)=e^{\pi\imath z}f\Big(z,\frac{1+\tau}{2}\Big)\,,\quad
  f_{3}(z)=f\Big(z,\frac{1}{2}\Big)\,.
  }
  \end{array}
   \eq
   Also, for distinct $\al,\be,\ga$
  \beq\label{A29}
  \begin{array}{c}
    \displaystyle{
  \varphi_\alpha(z)\varphi_\beta(z)=\varphi_\gamma(z)E_{1}(z)-f_\gamma(z)= - \partial_z \varphi_\gamma(z)\,.
  }
  \end{array}
   \eq
 The Appendix from \cite{Z11} can be also useful.

\subsection*{Acknowledgments}
 The work
was supported in part by RFBR grant 18-01-00273.
The research of A. Zotov was also supported in part by the HSE University Basic Research Program, Russian Academic Excellence Project '5-100' and by the Young Russian Mathematics award.

\begin{small}
 
\end{small}

\end{document}